\documentclass{aa}

\usepackage{natbib}
\usepackage{psfig}

\begin{document}

\title{Young isolated neutron stars from the Gould Belt}

\titlerunning{Young isolated neutron stars}

\author{S.B. Popov
            \inst{1,4}
             \and
M. Colpi\inst{2}
               \and
 M.E. Prokhorov\inst{1}
               \and
A. Treves\inst{3}
                \and
R. Turolla\inst{4}
         }

   \offprints{S. Popov}

   \institute{Sternberg Astronomical Institute,
Universitetski pr. 13, 119992 Moscow, Russia\\
\email{polar@sai.msu.ru; mike@sai.msu.ru}
            \and
Universit\`a di Milano-Bicocca,
Dipartimento di Fisica, 
Piazza della Scienza 3, 20126, Milano, Italy\\
\email{Monica.Colpi@mib.infn.it}
              \and
Universit\`a dell'Insubria,
Dipartimento di Scienze, 
via Vallegio 11, 22100, Como, Italy\\
\email{treves@mib.infn.it}
             \and
Universit\`a di Padova, Dipartimento di Fisica, 
via Marzolo 8, 35131, Padova, Italy\\
\email{turolla@pd.infn.it; popov@pd.infn.it}
}

   \date{}

\abstract{The origin of the local population of young, 
cooling neutron stars is
investigated with a population synthesis model 
taking into account the contribution of neutron stars born in the
Gould Belt, in addition to those originating in the Galactic
disk. We estimate their emission in the soft X-ray band as a
function of distance and age and construct the Log~N~--~Log~S
distribution. It is shown that the inclusion
of neutron stars from the Gould Belt provides a good fit to the
observed Log~N~--~Log~S distribution. As the Sun is situated
inside the Gould Belt, one can naturally explain the comparative
local overabundance of massive progenitors and can remove the
difficulty of the deficit of relatively bright ($\ga 0.1$ ROSAT 
PSPC cts~s$^{-1}$) cooling neutron stars previously reported from 
models where only
neutron stars from the Galactic disk were accounted for.
\keywords{stars: neutron  -- stars: evolution -- stars:
statistics -- X-ray: stars}}

\maketitle

\section{Introduction}

Observations of isolated neutron stars (INSs) are important for
gaining deeper insight on their structure and thermal evolution and 
ultimately might prove decisive in
unveiling the physical properties of matter at ultra-high
densities. Up to a decade ago, the only known INSs were active
radiopulsars, with the addition of the $\gamma$-ray pulsar
Geminga. Despite the fact that X-ray emission from some radio-pulsars
was already detected by {\it Einstein}, it was in the '90s
that ROSAT, thanks to its sensitivity in the  0.1-2 keV band,
gave a clearer picture of the faint X-ray emission produced by the
cooling surface of the closest INSs. ROSAT, supplemented by more
recent observations by Chandra and XMM-Newton, revealed a variety
of types of behavior in the X-ray emission from INSs and 
its relation with radio activity.

In particular, a substantial contribution of ROSAT to this field 
has been the
discovery of a group of seven radio-quiet, close-by,
thermally-emitting INSs, dubbed ROSAT INSs (RINSs) and
sometimes referred to as the ``magnificent seven''
(see for recent references \citealt{pasp}, \citealt{zam}, \citealt{hz}; 
see also \citealt{pp02}). The
nature of these sources has been controversial and two
interpretations were proposed, either as conventional
middle-aged cooling NSs or as very old NSs accreting the
interstellar gas (see again \citealt{pasp}). Although no compelling
evidence has been brought in favor of either picture as yet, it
is now generally believed that at least two of these objects (the
brightest ones, RX J1856-3754 and RX J0720.4-3125, e.g. \citealt{br}) 
are indeed relatively young cooling NSs.

Since the seven RINSs have remarkably similar observed properties, it is
quite natural to assume that they belong to the same class, i.e. all
of them are close-by, cooling NSs. This, however, poses a major
problem. A useful and standard way to study a population of sources
is to compute the Log~N~--~Log~S distribution starting from a model,
and then compare it with observations. This has been done for RINSs
by \cite{nt99} and \cite{p00b}. The main conclusion of 
the latter investigation
is that the typical spatial density of radio pulsars in our Galaxy is too
low to explain RINSs. In other words, the assumption that RINSs and
ordinary radio-pulsars derive from the same parent population underpredicts
the number of observed ``coolers'', i.e. NSs which are hot
and close enough for their thermal emission to be detected in X-rays.

An  obvious solution is to invoke a local (both in space and
time) overabundance of NSs with respect to those originating in
the Galactic disk and seen now as radiopulsars. The main goal
of this paper is to investigate the possible origin of these
objects. Here we suggest that the likely birthplace for many of
the INSs in the Solar proximity is the Gould Belt (see 
\citealt{p02} for a preliminary discussion). The Gould Belt
is a compound of young stellar associations extending in a ring
tilted at about $20^\circ$ from the Galactic plane, roughly
centered at the position of the Sun and with a radius of $\sim
1$~kpc. Originally discovered in the middle of the 19th century,
the Belt was studied in detail by Benjamin Gould (see
\citealt{StothersFrogel74} for a historical overview). Most of
the known close-by young star associations (in Ophiuchus, Orion,
Perseus, Scorpius etc.) belong to the Belt. Since the Gould Belt
is a relatively young stellar system it contains a comparatively
large fraction of massive stars. As the Solar proximity is
embedded within the Belt, it is likely to have harbored quite a number
of massive young stars which are the progenitors of NSs.

In Sect.~\ref{model} we perform 
a population synthesis of close-by young INSs,
assuming that they are born both in the Gould Belt and in the Galactic disk
with an assigned mass spectrum.
The dynamical evolution of the young NSs population in the Galactic potential
is then followed together with the thermal evolution. This allows us
to compute the present surface temperature and spatial distributions of young,
close-by NSs and finally their X-ray count rate, once a model for interstellar
absorption has been prescribed. This is used to compute the Log~N~--~Log~S
distribution which is then compared with observations in Sect.~\ref{results}.
We refer to ROSAT PSPC in evaluating the X-ray count rate and in confronting
our model with data. Even if Chandra and XMM-Newton
provide much better observations of individual sources and small fields, ROSAT
all-sky data are still the most complete by far in soft X-rays. Discussion and
conclusions follow in Sect.~\ref{discuss}.

\section{The Model}
\label{model}

In this section we describe the method we used to compute the local
distribution of young, cooling neutron stars. The main ingredients of our
model are: the spatial distribution of NS progenitors,
the NS formation rate, the NS cooling history, and the interstellar medium
(ISM) distribution (the latter determines the interstellar absorption and
hence affects the X-ray count rate). Even if we are concerned with a young
population ($\tau\la 10^6$ yrs), its dynamical evolution
can not be neglected if the number of sources in a limited volume (size
$\la 1$~kpc) has to be assessed. In fact, a NS with a typical velocity
$\sim 300$~km~s$^{-1}$ travels a distance $\sim 300$~pc in its lifetime
as a cooler.
While not much on a Galactic scale, such a displacement is 
non-negligible in evaluating the cooler population in the Solar
proximity. For this reason we account for  the dynamical evolution
of NSs in the Galactic potential in our model.

Our calculation proceeds in three steps. First, a spatial distribution
of the progenitors is selected and the ratio of NSs born in the
Galactic plane to those originating from the Gould Belt is fixed, together
with the birth rates (Sect.~\ref{spadistr}). Second, the dynamical
evolution is followed assuming that NSs at birth receive a kick velocity
drawn from a prescribed velocity distribution (see Sect.~\ref{evol}).
Finally, we derive the X-ray flux as a function of age and position
on the basis of an updated set of cooling curves (Sect.~\ref{cool}),
and translate this in ROSAT count rate for an assumed model of the
interstellar absorption (Sect.~\ref{ismabs}).

We varied the parameters of our model (NS formation rate, NSs mass
spectrum, time of calculation, time step, spatial distribution).
In the following subsections we focus on those
parameters corresponding to the results presented in Sect.~3.

\subsection{Initial spatial distribution}
\label{spadistr}

\begin{figure}
\vbox{\psfig{figure=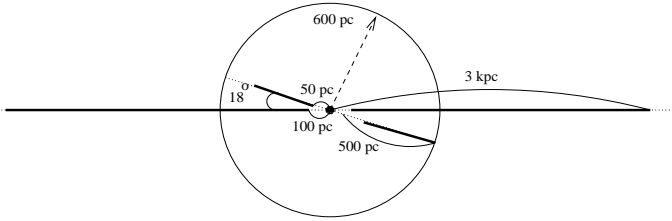,width=9.0cm}}
\caption[]{
$R-z$ projection of the initial spatial distribution of NSs. The heavy
lines mark the Gould Belt and the closer ($<3$ kpc) part of the Galactic
disk where NSs are assumed to be born. Note the two regions devoid 
of NS formation, in the Galactic disk with a
radius of 100 pc around the Sun, and in the Belt with a 
radius 150 pc around its center. The 0.6 kpc sphere centered at
the Sun is also shown for clarity.
}
\label{fig:scetch}
\end{figure}

In our picture NSs are born in the Galactic disk and in the Gould Belt
with a constant rate over the entire calculation (see Fig.~\ref{fig:scetch}
for a schematic view). This is a
reasonable approximation since we are
dealing with young objects, less than a few Myrs old.
The Gould Belt is modeled as a very thin disk of 500 pc in radius and
inclination to the Galactic plane of 18$^\circ$ with its center
situated at 100 pc from the Sun in the Galactic anticenter
direction. The central region of the Gould Belt is devoid of
massive stars (see \citealt{pop} for detailed description of the
Gould Belt, and \citealt{torra} for a shorter one). To mimic this,
we assume that no NS is born in the central region of 
the Belt up to a distance of 150 pc from its center.

We assume that the birthrate per unit area is independent of position,
i.e. it is constant both in the Belt and in the Galactic
disk. The two (constant) values are different, the one referring to the
Belt being larger.
To derive the NS birthrate, we rely upon direct counting of massive
stars which are doomed to end in a supernova event. This gives
$2.9 \times 10^{-11}$ massive progenitors per yr per pc$^2$ 
(\citealt{tammann})
which implies a NS birthrate of 30 NSs~Myr$^{-1}$ in 0.6 kpc around the Sun,
the region where data are available. 
In relating the NS and supernova birthrates we assumed that
core collapse produces a neutron star in 90\% of the cases. 
The above value
gives a more reliable estimate of the local NS birthrate, at present, 
as compared
with radio-pulsar and average supernova statistics.

 To get the separate contributions to the NS birthrate from the Belt
and from the disk for the region inside 0.6 kpc we proceed as follows.
In one Myr, out of a total of thirty, twenty NSs are born in the Gould Belt
according to the estimate of 20--27 supernova events given by \cite{g2000}.
The remainder is  uniformly distributed in the Galactic disk in a 
ring extending from 100 to 600 pc, to account for the known fact that
the closest Solar neighbourhood is underpopulated by massive stars 
(e.g. \citealt{maiz}).
This is in agreement with the results of
\cite{torra}, who estimate that about
2/3 of the massive stars inside 0.6 kpc belong to the Gould Belt,
and the rest to the Galactic disk.
 Direct counting of massive progenitors within 0.6 kpc from the Sun
accounts for all NSs originating in the Belt. However, the disk
clearly extends well beyond 0.6 kpc from the Sun. Although NSs
born far away are very dim their contribution may become
significant at low fluxes ($\la 0.1 \ {\rm cts}\, {\rm s}^{-1}$).
For this reason we decided to include also NSs born in the Galactic disk 
from 0.6 up to 3~kpc from the Sun. The NS birthrate per unit area
in this region is assumed to be again constant and coincident
with that in the disk inside 
0.6 kpc.
With this choice we find an agreement with existing data on
the supernova rate in a region of 1 kpc around the Sun
(\citealt{tammann}).

 In summary, we adopt a uniform birthrate of 20 NS Myr$^{-1}$ in 
the Gould Belt and 250 NSs~Myr$^{-1}$ in the Galactic plane up to a 
limiting distance of 3 kpc from the Sun. 
Our approach underestimates the number of distant ($\ga 1$ kpc) newborn NSs 
in the direction of the Galactic center because the rate of supernova events
is likely to increase there. However, 
interstellar
absorption towards the Galactic center is very high in the Galactic plane
(see Sect.~\ref{ismabs}), and this hinders the detection 
of sources at low Galactic latitude.
We note that the local overabundance of young NSs is probably due not only
to the enhanced number of massive stars harbored in the Belt but reflects also
the fact that the star formation history in the Belt favors an enhanced
supernova rate in the present epoch.

\subsection{Dynamical evolution}
\label{evol}

A detailed description of the evolutionary code may be found in
\cite{p00a, p00b}.
Typically we calculate about a
thousand evolutionary tracks in each run (up to $\sim 10^4$ per
production run) and then normalize our results to the actual
number of NSs born in the volume ($\sim 1000$) during the 4.25 Myrs time
interval, as discussed above.
An evolutionary track is calculated for specified initial position
and velocity of a NS. All tracks are used over their whole duration
and are applied to NSs of different masses,
i.e. for each track we have several different
cooling histories, the number of which depends on the number of different
masses used in the run.
As in Popov et al. (2000a,b)\nocite{p00a}\nocite{p00b},
each track actually represents a population of NSs of different masses
continuously born at the specified location with a prescribed spatial
velocity distribution, during the entire time interval of the calculation.

NSs are assumed to be born with a Maxwellian kick velocity distribution
with an average of $\sim 225$ km s$^{-1}$. Varying this parameter
does not change our results significantly because we are probing young
objects. The Galactic potential is assumed to be axisymmetric, resulting from
the sum of three contributions: the
disk, bulge and halo (see \citealt{pac}). The Sun is
in the Galactic plane at a distance 8.5 kpc from the Galactic Center.

\subsection{Cooling curves and flux calculation}
\label{cool}

\begin{figure}
\vbox{\psfig{figure=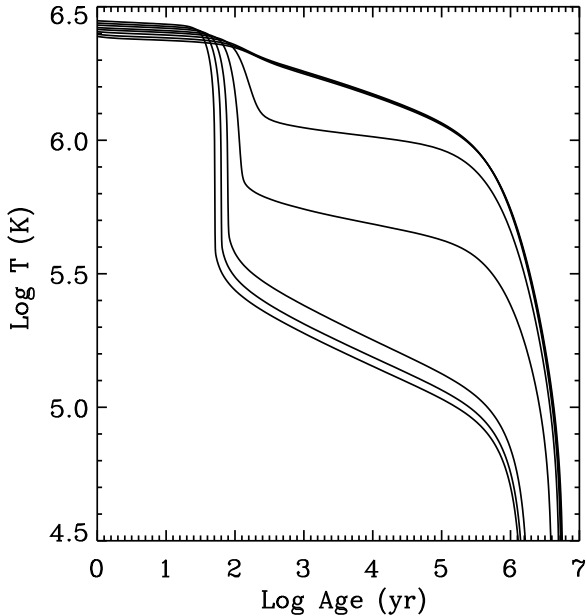,width=9.0cm}}
\caption[]{
Cooling curves for different NS masses (\citealt{kam}).
Curves from top to bottom correspond to masses
1.1--1.8~$M_{\odot}$ (step $0.1M_\odot$).
Here, and everywhere in the text, the
temperature is the red-shifted surface  temperature
(i.e. that observed at infinity).}
\label{fig:cool}
\end{figure}

To calculate the cooling of NSs we use the results of
the St.~Petersburg group (see \citealt{kam},
and the review by \citealt{yak}).
Cooling curves are for NSs of masses from $1.1\, M_{\odot}$ to
$1.8\, M_{\odot}$ with a step of $0.1\, M_{\odot}$ (see
Fig.~\ref{fig:cool}).
Curves take into account all important processes of neutrino emission.
The equation of state (EOS) used in \cite{kam}
was introduced by \cite{prak}. More precisely,
it corresponds to Model I of Prakash et al.
for symmetry energy and compression modulus of saturated
nuclear matter $K =240$~MeV. The maximum NS mass in this model
is 1.977 $M_{\odot}$. Since we base our calculation on a definite EOS,
the star radius $R$ is known once the mass $M$ is fixed. For
$1.1M_\odot <M< 1.8M_\odot$ it is $12.2\, {\rm km}<R< 13.2\, {\rm km}$,
where $R$ is circumferential radius.
The EOS used here corresponds to matter composed of neutrons, protons, 
and electrons (no muons, hyperons, and no exotic particles).
Neutron superfluidity was ignored.

We start to compute the emitted flux when a NS has an age of 10000 yrs.
This choice is motivated by the fact that the Vela pulsar (the youngest
close-by thermally emitting NS) is slightly older than that
and also because the NS birth rate used in our calculation corresponds to
an event every $\sim$ 10~000 yrs in the $\sim 1.7$ kpc region around the Sun.
For each NS the calculation is stopped when the temperature drops below
$10^5$ K. This happens at an age of 4.25 Myrs for the lightest NSs
($M=1.1\, M_{\odot}$) or less for more massive stars.
NSs that cold could have been detected by ROSAT PSPC within a distance
of only 10 pc with a count rate $\sim 5\times 10^{-3}$ cts s$^{-1}$.
We assume that emission comes from the entire star surface. This
appears reasonable in the light of the relatively low pulsed fractions
($\la 15$-20\%) detected so far in the majority of isolated NSs
(see \citealt{hz}). We do not take into account any reprocessing of
the blackbody surface emission in the NS
atmosphere.

 NSs are expected to have a mass spectrum (see discussion in \citealt{whw02}).
In the calculations we present here
NSs are taken to have a flat mass spectrum
between 1.1 and $1.8M_\odot$.
The mass spectrum (and clearly the mass of each star) 
are assumed to be constant during the calculations.
With the above values (maximum age 4.25 Myrs and time step 10000 yrs)
each evolutionary track represents $425\times 8$ NSs of eight
different masses born with a time step $10^4$ yrs at the same place with
the same initial velocity in a period 4.25-0.01 Myrs ago.

As one can see from Fig.~\ref{fig:cool}, the main
contribution to X-ray bright sources comes from NSs with masses
$<1.5 \, M_{\odot}$. Young isolated NSs which are observed in several
SN remnants as compact X-ray sources
also should be relatively low-mass objects (\citealt{kam}).
Observations of binary radio pulsars suggest that NSs masses are strongly
peaked around 1.35 $M_{\odot}$ (\citealt{tho99}).
We repeated our calculations using a similar distribution and found
that it produces similar results, although the number
of observable isolated NSs increases  by $\sim$ 30 \%.
 Calculations which take into account the realistic mass function of massive
progenitors in the Solar vicinity will be the subject of a future paper.

\subsection{ISM and absorption}
\label{ismabs}

Since young cooling NSs are expected to emit most of their
luminosity at UV/soft X-ray energies ($\sim 20-200$~eV,
corresponding to temperatures  $\sim 10^5$--$10^6$~K),
interstellar absorption plays a crucial role with respect to their
observability. Any attempt to derive the number of observable
cooling isolated NSs using the unabsorbed flux would result in a
substantial overestimate.

For the ISM distribution we use the same prescription
as in \cite{pp98}.
The Local Bubble is modeled as a sphere with a radius of
140 pc and ISM density of 0.1 cm$^{-3}$.
Typical column densities for sources inside the calculated volume
are in the range $N_{\rm H}\sim 10^{19}$--$10^{21}$~cm$^2$.
After the column density is evaluated for a current NS position,
we calculate the unabsorbed flux corresponding to the
temperature, radius
of the NS and its distance from the Sun, and apply the standard
procedure to derive the ROSAT PSPC count rate.
Outside the Local Bubble in directions close to the Galactic plane,
absorption starts to play a crucial role. That is why regions
closer to the Galactic center ($<7$ kpc), where the NS formation
rate should be higher than in the Galactic disk in the Solar vicinity,
cannot add many sources to our sample.
At $N_{\rm H}= 3\times 10^{21}$~cm$^{-2}$ even
very young and hot low-mass NSs with $kT\approx 0.1 $ keV cannot have
been  detected at a distance 1 kpc in the ROSAT Bright Survey (RBS)
at the threshold of $\sim  0.2$~ cts~s$^{-1}$.

Except for the Local Bubble (and even that in a quite simplified
way, see \citealt{sfeir} for a more complete description),
our model does not take into account
small-scale irregularities in the ISM
distribution. They can be important if one makes an attempt to produce
a realistic map of RINSs distribution on the celestial sphere,
but in the case  of the all-sky averaged Log~N~--~Log~S
distribution our approximation is adequate.

\section{Results}
\label{results}

\begin{figure}
\vbox{\psfig{figure=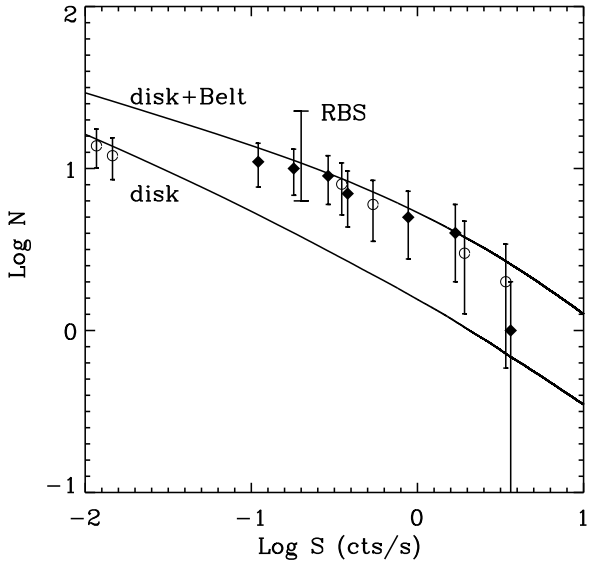,width=9.0cm}}
\caption[]{All-sky Log N - Log S distribution:
filled diamonds are the seven RINSs and
open circles Geminga, the ``three musketeers'', PSR 1929+10 and 3EG
J1835+5918. We also show the RBS limit (Schwope et al. 1999).
Upper curve: NSs born in the Gould Belt and in the Galactic disk
(total birth rate 270 NS~Myr$^{-1}$).
Lower curve: disk only (birth rate 250 NS~Myr$^{-1}$).}
\label{fig:lnls}
\end{figure}

As discussed in the previous section, the dynamical evolution of young
neutron stars, together with the calculation of their X-ray flux, allows
us to derive the Log N -- Log S for these sources. In particular, having
fixed the number of NSs originating in the Gould Belt and in the Galactic
disk, we can compute the separate contributions of these two sub-populations
to the total Log N -- Log S. Our main results are presented 
in Fig.~\ref{fig:lnls} where the 
total (disk+Belt) and the disk-only distributions
are compared with observations. All curves refer to the whole sky, i.e.
the angular coverage used here is the entire solid angle.
The observed Log N -- Log S has been derived from ROSAT data of
the seven RINSs and six other close-by
young isolated NSs (see Table 1). Contrary to RINSs, the latter
sources exhibit a composite spectrum with a non-thermal high-energy tail
superimposed on the thermal component. The total count rate has been used
in this case, but we stress that the non-thermal contribution is sizeable
only for the Vela pulsar and PSR 1929+10. 
Poissonian errors are assumed in both observations and simulations. 
The computed curves were generated with a 
sufficiently large number of stars to reduce the statistical noise
which is particularly severe at large flux (see \S \ref{evol}).
Statistical errors are shown for the observational points only, being too
small to be appreciated in the plot of the simulated distributions.
Uncertainties arising from the assumed model parameters are more 
difficult to assess and could introduce non-negligible deviations. Test
runs with different choices of key parameters show that differences
are of the order of 50\%. 
Symbols which show observational points (filled diamonds or open circles)
 correspond to the type of the faintest object (RINS or not)
which contributes to the total number at the specified count rate.

Previous investigations have convincingly shown
that cooling NSs with the same spatial density of ordinary
radio-pulsars cannot
explain the observed Log~N~--~Log~S distribution for RINSs
(\citealt{nt99}, \citealt{p00b}). In particular, it was stressed
that an additional contribution to the local population of young
cooling NSs should be invoked to explain the relatively large
number of bright sources at fluxes $\ga 0.1$ cts~s$^{-1}$.
The Gould Belt is a natural candidate to provide  the missing
NSs in the Solar proximity.
The population synthesis calculations presented here strongly
support this claim and account not only for the observed distribution
of RINSs but also for that of other young, close-by isolated NSs
observed by ROSAT.

As can be seen from Fig.~\ref{fig:lnls}, the Gould Belt provides the
major contribution to the local population of young cooling NSs and
the theoretical prediction gives a very good fit to the data, 
within statistical errors. The
contribution of NSs born in the Galactic disk is not very important
at relatively large fluxes (it leaves rooms for only a few sources
with count rate $\ga$ 0.1
cts~s$^{-1}$), but, as expected, becomes dominant at lower fluxes
($\la$~0.01~cts~s$^{-1}$) where far away stars contribute most.
Fig.~\ref{fig:lnls} clearly shows that the curve referring to the Galactic
disk alone is always below the observed points with high statistical 
significance
and below the RBS limit. 
The worsening of the agreement at low 
fluxes ($\la$~0.05~cts~s$^{-1}$) is to be attributed to the 
incompleteness of the X-ray sample.
The deficit of very bright objects ($\ga 1$ cts~s$^{-1}$) may be attributed 
to chance statistics. 

Our calculations show that there can be about $10-50$ unidentified
isolated NSs in the ROSAT All-Sky Survey (RASS) at a limiting
flux of $\ga 0.015$ cts s$^{-1}$ depending on parameters of the model.
 The number of INSs in the RASS/BSC has been recently
investigated by 
\cite{rut2003}, who found that at most 67 sources could have been detected
at the 0.05 cts s$^{-1}$ level and have escaped 
identification. Our results are well within this limit.
Also there may be a few unidentified RINSs at fluxes $\ga 0.1$ cts s$^{-1}$
at low Galactic latitudes (see also \citealt{sch}). 
Most sources should be observed
at $\pm$ 20$^\circ$ from the Galactic plane towards the directions
of lower absorption.
Some of them may be Geminga-like objects with counterparts among unidentified 
gamma-ray sources (also connected with the Gould Belt, see \citealt{g2000}).
Identification of these objects
may prove important for constraining cooling models and the NS mass spectrum.

Absorption, the flat geometry of NS initial distribution and
the finite extension of the Gould Belt naturally explain the
very flat (slope flatter than -1) Log~N -- Log~S curves in Fig.~\ref{fig:lnls}.
Our model contains some degrees of freedom (e.g. the NS mass spectrum, 
details of the formation rate and
surface emission, a simple blackbody was used here) that can be
varied. We nevertheless
believe that even this simple picture  (albeit based on realistic assumptions)
gives a quite satisfactory explanation
of the observed properties of isolated NSs in the Solar proximity.
Preliminary calculations which take into account realistic mass
spectrum of NS progenitors, atmospheric effects and different
spatial and velocity distributions are in progress and show no 
qualitative changes.

\begin{table*}[t]
\caption{Local ($r<1$ kpc) population of young (age $<4.25$ Myrs)
isolated neutron stars \label{table}
}
\begin{tabular}{|l||c|c|c|c|c|c|}
\hline
\hline
 & & & & & & \\
Source name & Period, & Count rate, & $\dot P$ & Distance & Age$^a$ & Ref. \\
            &   s     & ROSAT cts s$^{-1}$ & $10^{-15}$ ss$^{-1}$& kpc   & Myrs     &      \\
 & & & & & & \\
\hline
 & & & & & & \\
RX J185635-3754            &  ---  & 3.64  &  ---&0.117$^d$&$\sim0.5$& [1,2]\\
RX J0720.4-3125                 & 8.37 & 1.69  &$\sim 30-60$& ---&---& [1,3]\\
1RXS J130848.6+212708 (RBS 1223) & 10.3 & 0.29  & $< 10^{4}$?&---&---&[1,4] \\
RX J1605.3+3249 (RBS 1556)       &  ---  & 0.88  & --- & --- & --- & [1]\\
RX J0806.4-4123                 &  11.37  & 0.38  & --- &---&--- &  [1,5]\\
RX J0420.0-5022                 &  22.7   & 0.11  & --- &---   &--- & [1]\\
1RXS J214303.7+065419 (RBS 1774) &  ---    & 0.18  & --- & ---  &--- & [6]\\
 & & & & & & \\
\hline
 & & & & & & \\
PSR B0633+17 (Geminga)           & 0.237 & 0.54$^c$ &10.97&0.16$^d$&0.34& [7]\\
RX J1836.2+5925 (3EG J1835+5918) & ---   & 0.015    & --- & ---  &  --- & [8]\\
 & & & & & & \\
\hline
 & & & & & & \\
PSR B0833-45   (Vela) & 0.089 & 3.4$^c$  & 124.88 & 0.294$^d$ & 0.01&[7,9,10]\\
PSR B0656+14          & 0.385 & 1.92$^c$ &  55.01 & 0.762$^e$ & 0.11 &[7,10]\\
PSR B1055-52          & 0.197 & 0.35$^c$ &   5.83 & $\sim 1^b$ & 0.54&[7,10]\\
PSR B1929+10          & 0.227 & 0.012$^c$& 1.16 &  0.33$^d$  & 3.1& [7,10]\\
 & & & & & & \\
\hline
 & & & & & & \\
PSR J0056+4756  & 0.472 & --- & 3.57 &  0.998$^e$  & 2.1& [10]\\
PSR J0454+5543  & 0.341 & --- & 2.37 &  0.793$^e$  & 2.3& [10]\\
PSR J1918+1541  & 0.371 & --- & 2.54 &  0.684$^e$  & 2.3& [10]\\
PSR J2048-1616  & 1.962 & --- & 10.96&  0.639$^e$  & 2.8& [10]\\
PSR J1848-1952  & 4.308 & --- & 23.31&  0.956$^e$  & 2.9& [10]\\
PSR J0837+0610  & 1.274 & --- & 6.8  &  0.722$^e$  & 3.0& [10]\\
PSR J1908+0734  & 0.212 & --- & 0.82 &  0.584$^e$  & 4.1& [10]\\
 & & & & & & \\
\hline
\hline
\end{tabular}
\begin{tabular}{l}
$^a$) Radio-pulsar ages are estimated as $P/(2\dot P)$,\\
      for RX J1856-3754 the age estimate comes from kinematical
      considerations (\citealt{wl02}).\\
$^b$) Distance to PSR 1055-52 is uncertain ($\sim$ 0.9-1.5 kpc)\\
$^c$) Total count rate (blackbody + non-thermal)\\
$^d$) Distances from parallactic measurements\\
$^e$) Distances from the dispersion measure\\
(1)  \cite{pasp},
(2) \cite{kaplan},
(3) \cite{zane},\\
(4) \cite{hamb}, 
\cite{h2000},
(5) \cite{hz},\\
(6) \cite{zam},
(7) \cite{bt97},
(8) \cite{mh},\\
(9) \cite{pav},
(10) ATNF Pulsar Catalogue,
(http://wwwatnf.atnf.csiro.au/research/pulsar/catalogue/)\\
 \\
\hline
\end{tabular}
\end{table*}

\section{Discussion and conclusions}
\label{discuss}

At  present about 20 nearby (distance $<1$ kpc), young (age $<4.25$ Myrs)
isolated NSs are known (see Table \ref{table}). The local NS population
includes objects with different properties: radio-quiet, thermally-emitting
NSs (the seven RINSs), Geminga and
the Geminga-like object 3EG J1835+5918 (these are probably active
pulsars with radiobeams missing
the Earth), radio-pulsars with observed thermal X-ray emission
(PSR 1929+10 and the ``three musketeers'': the Vela pulsar, PSR 0656+14,
PSR 1055-52) and seven other pulsars which are not
detected in X-rays. The latter objects are relatively old in comparison
with the others and lie further away, as  can be seen from Table \ref{table}
where the wide
gap in the estimated age of PSR 1055-52 and PSR J0056+4756 is apparent.
For ages of a few Myrs, even very low-mass (and hence slow-cooling) NSs
are too cold by now to have been detected with ROSAT at distances
$\ga 0.5$~kpc. The case of PSR 1929+10 is intermediate since it is
$\sim 3$~Myrs old but relatively close ($\sim 300$~pc) and its X-ray
emission is mainly non-thermal.

As  has been discussed by \cite{nt99} and \cite{p00b}, the number
of X-ray bright INSs is too large to be explained in terms
of the average density of radio-pulsars in the Solar neighborhood.
This motivated the suggestion that a sizeable fraction of NSs
never become active radio emitters, as suggested at the same time
by considerations on young NSs in supernova remnants
\citep{gv} albeit the origin of the local NS population was left open.
In this paper we have shown that thermally-emitting
INSs are naturally explained
as cooling NSs born in the Gould Belt. The relatively high
local spatial density of young NSs is then due to the large number
of massive progenitors in the young stellar associations which
constitute the Gould Belt. Our analysis lends support to the idea
that RINSs (the ``magnificent seven'') represent the slowly cooling
NSs and hence a clean sample of NSs with $M\la 1.3M_\odot$ 
(\citealt{kam}).

The computed Log~N~--Log~S distribution for X-ray thermally-emitting
INSs born in the Gould Belt and (to a lesser extent) in the Galactic
disk accounts for all bright INSs in the Solar vicinity
(see Fig.~\ref{fig:lnls}) and leaves room for at most
1-2 undetected sources above $\sim 1$ cts s$^{-1}$.
 The absence of sources brighter than RX J1856 (Log S$ > 0.56$) 
is  consistent with our calculations at the 2~$\sigma$ level.
Having neglected very young NSs (age $<10000$ yrs) does not change our 
conclusions significantly.
Only young stars with $M<1.35\,M_{\odot}$ (about 1/2 of their total
number) have temperatures about $2\times 10^6$~K (see Fig.~2), which 
corresponds to a luminosity $\sim 10^{34}$ erg~s$^{-1}$.  
With our NS formation rate we 
expect to have one such  NS within a distance of 
$\sim 2.5$ kpc. About half of the NSs born in that region should originate
inside 1.7 kpc. Placing the source at this distance its unabsorbed 
flux is $\sim 3\times 10^{-11}$ erg~cm$^{-2}$~s$^{-1}$. With a typical 
column density of $10^{21}$~cm$^{-2}$, Web-PIMMS gives 
$\sim 1.3$ cts~s$^{-1}$ for ROSAT PSPC.  
We can then conclude that stars younger than 10000
yrs can contribute only to fluxes $<1.3$~cts~s$^{-1}$.
This translates into a $\sim 20\%$ difference in the Log N - Log S curve
shown in Fig.~\ref{fig:lnls} in the worst case. Typical uncertainties of 
the present model are at about the same level.

The local ($\la 1$~kpc) INS population was estimated to comprise total
about one hundred objects younger than $\sim $4 Myr, taking into
account that some NSs born inside 1 kpc can leave the local space
in their lifetime. These NSs are not detected as radio pulsars,
but tens of them could be identified
in ROSAT surveys as dim sources. The beaming effect can be responsible
only for part of these young NSs to be radiosilent. According to
\cite{tm98} in fact, about 50-70\%
of young pulsars are not visible from the Earth (see also 
\citealt{bra99}). In our model this would
imply the presence of at least $\sim 30$ active pulsars with an age
$\la 4$~Myrs in the Solar vicinity. However, only about 1/3 of such
pulsars are
observed (see again Table \ref{table}).  
A bimodal velocity distribution, with
the high charcteristic velocity $\ga 500$ km~s$^{-1}$  
(\citealt{acc02}), would reduce the discrepancy since a number of young
pulsars leave the local volume considered here. However, even barring 
observational bias against low-luminosity radio-pulsars, our model seems  
to support the argument by
\cite{gv}, that at least half of the observed young neutron stars follow an
evolutionary path quite distinct from that of the Crab pulsar.

\begin{acknowledgements}
We want to thank Dmitry Yakovlev for putting his cooling model to our
disposal and for his invaluable help with it. We are also indebted to
Vasily Beskin, Matteo Chieregato and Andrea Possenti for many
useful discussions. The work of SP was supported by the
Russian Foundation for Basic Research (RFBR) grant 02-02-06663 and by
RSCI. SP thanks the Universities of Insubria at Como and
Milano-Bicocca, where part of this investigation was carried out,
for hospitality.
The work of MC, AT and RT was partially supported by the
Italian Ministry for Education, University and Research (MIUR) under grant
COFIN-2000-MM02C71842.

\end{acknowledgements}

\end{document}